\documentclass[12pt,onecolumn,tworows]{IEEEtran}
\usepackage{amsfonts,amssymb,amsmath,bm}
\usepackage{graphicx,subfigure}
\usepackage{cite,color}
\usepackage{booktabs}
\usepackage{threeparttable}
\usepackage{mathtools}
\usepackage{verbatim}
\usepackage{cite}

\begin{document}
%
\title{Distributed Energy Efficient Cross-layer Optimization for Multihop MIMO Cognitive Radio Networks with Primary User Rate Protection}
\author{Weiqiang Xu,~\IEEEmembership{Senior Member,~IEEE},
 Wenchu Yuan,\\
Qingjiang Shi,~\IEEEmembership{Member,~IEEE},
Xiaodong~Wang,~\IEEEmembership{Fellow,~IEEE},
Yake Zhang
\thanks{Weiqiang Xu, Wenchu Yuan, Qingjiang Shi, and Yake Zhang are with School of Information Science \& Technology, Zhejiang Sci-Tech University, Hangzhou, 310018, P. R. China. Email: wqxu@zstu.edu.cn, wenchuy@zstu.edu.cn, shiqj@zstu.edu.cn, yakezhang@zstu.edu.cn}
\thanks{Xiaodong Wang is with the Department of Electrical Engineering, Columbia University, New York,
NY, 10027, USA. Email: wangx@ee.columbia.edu.}
}
\maketitle

\baselineskip 25pt
\begin{abstract}
\baselineskip 18pt
Due to the unique physical-layer characteristics associated with MIMO and cognitive radio (CR), the network performance is tightly coupled with mechanisms at the physical, link, network, and transport layers. In this paper, we 
 consider an energy-efficient cross-layer optimization problem in multihop MIMO CR networks. The objective is to balance the weighted network utility and weighted power consumption of SU sessions, with a minimum PU transmission rate constraint and SU power consumption constraints. However, this problem is highly challenging due to the nonconvex PU rate constraint. We propose a solution that features linearization-based alternative optimization method and a heuristic primal recovery method. We further develop a distributed algorithm to jointly optimize covariance matrix at each transmitting SU node, bandwidth allocation at each SU link, rate control at each session source and multihop/multi-path routing. Extensive simulation results demonstrate that the performance of the proposed distributed algorithm is close to that of the centralized algorithm, and the proposed framework provides an efficient way to significantly save power consumption, while achieving the network utility very close to that achieved with full power consumption.
\end{abstract}

%
\IEEEpeerreviewmaketitle
\newpage

\section{Introduction}

The cognitive radio (CR), with its built-in intelligence and cognitive capabilities,
can flexibly adapt its transmission or reception parameters, and
provides the means for unlicensed secondary users (SU)
to dynamically access the licensed spectrum held by primary users (PU) \cite{Haykin2005}.
Thus, as a promising technology to deal with the spectrum under-utilization problem, CR has quickly become the enabling technology for the next-generation wireless communications, and will be adopted by many emerging applications, e.g., smart grid communications, public safety, and medical applications\cite{WangGhosh}.



Research on effective spectrum sharing or spectrum
allocation has been extensive. For multi-user single-hop CR networks (CRN), a number of approaches have been proposed.
For example, \cite{ZhangSPM2010} provided a survey on dynamic resource allocation schemes with the interference temperature based spectrum-sharing model.
For the multihop networking problem with CRs, there is
a limited amount of work to date available in the literature \cite{Hou2008,Shiyi2008,Wang2011,GaoShi2011,YangSong2010,JSAC2013Routing,JSAC2013Duality,JSAC2013Delay,LiangSurvey}.
\cite{Hou2008} proposed a mathematical formulation to modeling spectrum sharing, sub-band division, scheduling, and flow routing in multi-hop CRN.
\cite{Shiyi2008}  developed a formal mathematical model for a joint per-node based power control, scheduling, and flow routing problem in multi-hop CRN. These two joint formulation are the mixed-integer non-linear program (MINLP) problem.
\cite{Wang2011} proposed a framework of joint spectrum allocation and power control to utilize the open spectrum bands in CRN with both interference temperature constraints and spectrum dynamics.
\cite{GaoShi2011} investigated a multicast communication problem in a multihop CRN. A scheduling and routing approach was proposed to minimize the network-wide resource utilization to support a set of multicast sessions.
\cite {YangSong2010} addressed the stochastic traffic engineering problem in multihop CRN. The challenges induced by the random behaviors of the primary users are addressed through the stochastic network utility maximization framework.
\cite{JSAC2013Routing} proposed a distributed routing algorithm to minimize the aggregate interference from the SUs to the PUs in CR mesh networks.
\cite{JSAC2013Duality} studied a utility maximization framework by adapting SINR assignment and transmit power subject to power budget constraints and additional interference temperature constraint.
\cite{JSAC2013Delay} investigated the problem of spectrum assignment and sharing to minimize the total delay of multiple concurrent flows in multi-hop CRN.
\cite{LiangSurvey} provided a survey on the state-of-the-art of research on physical, medium access, and routing layer issues in the design of CRN, including the multihop scenario.

In parallel to the development of CR, MIMO is a physical
layer technology that can provide many types of benefits through multiple antennas and advanced signal
processing.

The potential
network capacity gain with the use of MIMO depends on the
coordinated mechanisms at the physical, link, and network
layers\cite{Mietzner2006}. Thus, some works have addressed to exploit the benefits of MIMO from a cross-layer prospective in single and multihop MIMO ad hoc networks.
For example,
\cite{Kim2007} investigated the cross-layer optimization problem in
multihop
MIMO backhaul networks to maximize the fair throughput of the
access points.
\cite{Chu2008} proposed the cross-layer algorithms for
MIMO ad hoc networks to maximize the SINR.
\cite{LiuJSAC2008} considered the
problem of jointly optimizing power and bandwidth allocation
at each node and multihop multi-path routing in a MIMO
ad hoc network, and developed a two-step solution to this
cross-layer optimization problem.

In CR networks, some efforts to explore the cognitive MIMO radio mainly focus on the physical layer.
The game theoretical approach and optimization-based approach have been applied to explore the potential of MIMO CRN\cite{Scutari2008,Scutari2010,Bixio2010,RZhangSTSP2008,Nguyen2012,Kim2011IT,YingJunZhang2011,Ebrahim2011,JSAC2013Energy}.
For instance, \cite{JSAC2013Energy} proposed joint time scheduling and beamforming optimization to minimize SU's energy consumption while satisfying SU's rate requirements and PU interference constraints.
However, these efforts mainly focused on the resource optimization for single-hop MIMO CRN. Few effort has addressed how to take advantage of the MIMO techniques in the
context of multihop CRN through cross-layer design.
\cite{CunhaoGao2011} developed a tractable mathematical model for multihop MIMO
CRN and jointly optimized channel assignment in CRN and degree-of-freedom allocation in MIMO to maximize the throughput. However, the centralized method is difficult to be implemented in the realistic multihop MIMO CRN.

In this paper, we propose a new formulation to address an energy efficient cross-layer optimization problem in a multihop MIMO CRN. Our objective is to optimize the weighted sum of network utility and power consumption with a minimum PU transmission rate constraint and SU power consumption constraints.
We jointly optimize covariance matrix
at each transmitting SU node, bandwidth allocation
at each SU link, rate control at each session source and multihop/multi-path routing. However, such a joint formulation is usually highly nonconvex and difficult to solve due to the nonconvex MIMO link capacity constraint and the PU rate constraint. Although by exploiting the special structure we can easily design a distributed algorithm based on the popular dual-decomposition method, the dual-decomposition solution is not feasible to the primal problem due to the nonconvexity nature of the problem. Hence, in this paper, we attempt to design a distributed algorithm that can yield a feasible suboptimal solution. Our algorithm includes two steps. The first step is to obtain a reasonable bandwidth allocation while in the second step we recover a feasible solution. Both steps are based on the dual-decomposition method, which decomposes the joint optimization problem into the network-transport layer subproblems and the physical-link layer subproblems. The network-transport layer subproblems in both steps are convex and can be solved efficiently and distributively. We mainly focus on the nonconvex physical-link layer subproblem in the first step and propose an iterative algorithm based on linearization of the nonconvex constraint and alternating optimization. We prove the algorithm can converge monotonically. The performance of our proposed distributed optimization algorithm is examined by extensive simulations. By comparing with a performance upper bound offered by a centralized algorithm, we find that the proposed algorithm can provide a feasible solution with good performance, and the proposed formulation can provide an efficient mechanism to significantly save power consumption, while achieving the network utility very close to that achieved at full power consumption.
The key novelties and contributions of our work are summarized as follows:
\begin{itemize}
 \item Note that for the MIMO-link capacity function Eq. (2) used in \cite{LiuJSAC2008},
the background noise is assumed to be not related to the allocated bandwidth.
In this paper, we use a correlated but different MIMO-link capacity formula, where
the background noise is related to the allocated bandwidth.
Based on  a linearization-based alternative optimization method and a heuristic primal recovery method, we design a new distributed algorithm that can find a feasible solution with good performance for the non-convex optimization problem of multihop MIMO CRN, which is a generalization of the optimization problem in \cite{LiuJSAC2008}.

\item Up to now, MIMO CRN design has mainly been focusing on utility (e.g., sum-rate) maximization or the power minimization separately. However, achieving high transmission rate while consuming low power consumption has recently become a main concern. In this  paper, we extend traditional MIMO CRN design towards a framework of utility-power trade-off with power related objective and constraint. Specifically, we address the optimization objective of the weighted sum of network utility and power consumption and consider non-full power consumption constraint. We quantify the utility-power trade-off with some surprising numerical results. Thus, our formulation provides an efficient way to design a green cross-layer optimization scheme for multihop MIMO CRN.

\item So far, many existing works proposed various algorithms for MIMO CRN based on the interference power constraints for protecting PU. In contrast to the conventional interference power constraint approach, we protect the primary receiver by the direct constraint on the minimum PU transmission rate subject to the SU interference.
Although the non-convexity of the minimum PU rate constraint is introduced, we apply the method of Taylor series expansion to approximate this non-convex constraint and propose an iterative algorithm to solve the non-convex optimization problem.
                                                       \end{itemize}


The remainder of this paper is organized as follows. In Section II, we present the system model and problem formulation. In Section III, we solve our problem through the dual decomposition and the subgradient algorithm. In Section IV, we propose the linearization-based alternative optimization method. In Section V, we recover the primal solution. In Section VI, we provide the simulation results. Finally, we conclude the paper in Section VII.

The upper-case and lower-case boldface letters denote matrices and vectors, respectively. The conjugate transpose, Hermitian transpose, determinant are represented by ${( \cdot )^*}$, ${(\cdot)^H}$, $\left|\cdot  \right|$, respectively. The trace is denoted by $\text{Tr}( \cdot )$. We let ${{\bf{I}}}$ denote the identity matrix, the dimension is determined by the context. We define the operators $ \ge $ and $ \le $ for vectors and matrices as componentwise, while ${\mathbf{A}}\underset{\raise0.3em\hbox{$\smash{\scriptscriptstyle-}$}}{ \succ } {\mathbf{B}}$ means that ${\mathbf{A}}-{\mathbf{B}}$ is positive semidefinite.

\section{System Model and Problem Formulation}
We consider a multihop MIMO CRN, where multi-antenna SUs share the same spectrum with multi-antenna PUs.
We focus on the spectrum sharing model. In this model, from PU's perspective, SU is allowed to transmit as long as the
interference from SU does not degrade the quality of service
(QoS) of PU to an unacceptable level. From SU's perspective,
SU should control its transmit power properly in order to
achieve a reasonably high transmission rate without causing
too much interference to PU.
The SUs form a multihop FDMA MIMO network, which is represented by a directed graph, denoted as ${\mathcal{G} _s} = \{ {\mathcal{N} }_s,{\mathcal{L}}_s\}$, where ${\mathcal{N}}_s = \{ 1,2,\ldots,n_s,\ldots,N_s\} $ and ${\mathcal{L}}_s = \{ 1,2,\ldots,l_s,\ldots,L_s\}$ represent all SUs and all the possible SU MIMO links, respectively. The network is assumed to be always connected. Within such a multihop MIMO CRN, the packets from a source node will reach a destination
node through multihop/multipath.

\subsection{Application Scenario}
\begin{figure*}
\centering
\includegraphics[scale=0.8,bb= 15 540 595 822]{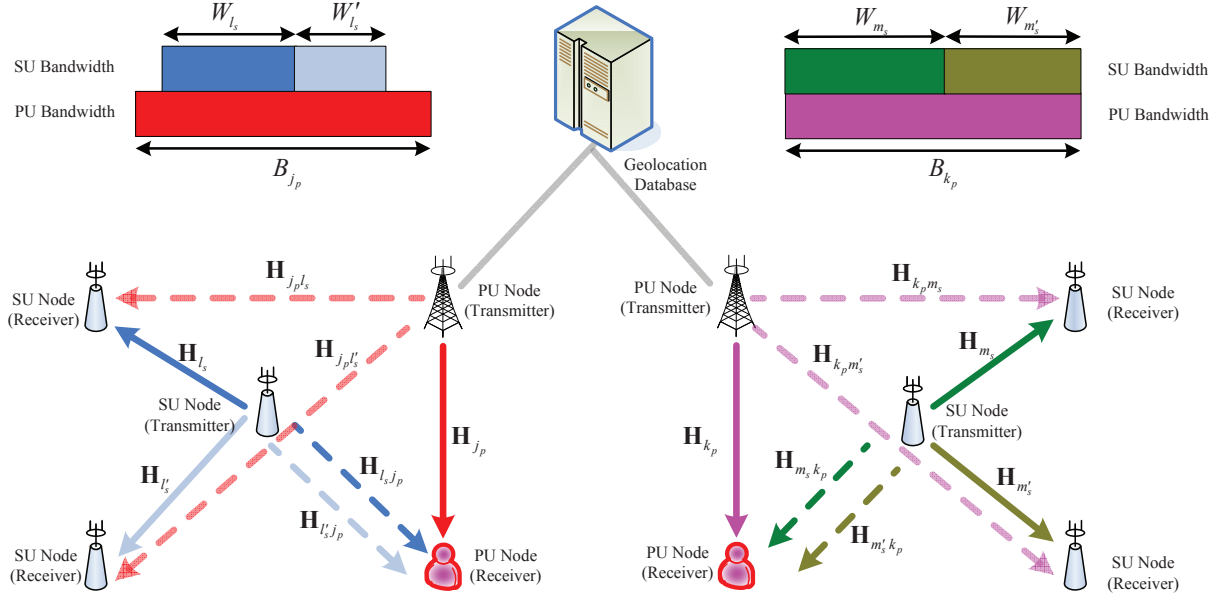}
\caption{An application scenario of multihop MIMO CRN operating over TVWS.}
\label{fig:TVWS_scenario}
\end{figure*}

Recently, a
series of emerging wireless access standards incorporated
CR technology over Digital TV white space (TVWS)\cite{TVoverview} .
Fig. \ref{fig:TVWS_scenario} gives an application scenario of multihop MIMO CRN operating over TVWS. The FCC ruling had opened up the possibility of designing wireless multihop networks where CR operates over TVWS \cite{FCC2012Database}. The most recent FCC ruling requires that TVWS devices must rely on a geolocation database to determine the spectrum availability \cite{Chen2013Database}\cite{Murty2012Database}. In such a database-assisted architecture, the primary licensed holders of TV spectrum provide the database with the up-to-date information including TV tower transmission parameters, TV receiver protection requirements, and etc..
Based on this information, the geolocation database implements a centralized spectrum allocation mechanism such that different SU nodes are assigned non-overlapping bandwidths. In this paper, we assume that the bandwidth of each SU node is given from the geolocation database. Our proposed bandwidth allocation scheme is to re-allocate/sub-divide the bandwidth of SU node among its all outgoing links.
Note that its incoming and outgoing links is assigned different frequency bands, such that  SU node can simultaneously transmit and receive signals, and
cause interference among each other.

\subsection{ Channel Model}
A MIMO SU $n_s \in {\mathcal{N} }_s$ includes possible multiple secondary outgoing links  ${l_s} \in \mathcal{O}\left( {{n_s}} \right)\subset{\mathcal{L}}_s$, where $\mathcal{O}({n_s})$ denotes the set of links that are outgoing from node ${n_s}$.
Without loss of
generality\footnote{Note that it is readily seen that we can easily generalize to a scenario where a SU can coexist with multiple PUs, and our proposed scheme also applies to this scenario with minor modification.}, we assume that a MIMO SU $n_s$ only shares a portion or the entire of the bandwidth with a MIMO PU link $j_p$.
Let $B_{{j_p}}$ denote the total bandwidth of PU link,
${W_{{l_s}}}$  the overlapping bandwidth of PU link ${j_p}$ and SU link ${l_s}$.
Furthermore, we assume that the transmit power of each PU link is distributed equally over frequency band, i.e., the power allocation of the PU link for different frequency band is proportional to the bandwidth. The transmission over the SU MIMO channel ${l_s}$ with ${T_{{l_s}}}$ transmit and ${R_{{l_s}}}$ receive antennas can be expressed as the following signal model:
\begin{equation}\label{1}
{{\mathbf{r}}_{{l_s}}} = {{\mathbf{H}}_{{l_s}}}{{\mathbf{t}}_{{l_s}}} + \sqrt{\frac{W_{l_s}}{B_{j_p}}}\mathbf{H}_{j_pl_s}\mathbf{t}_{j_p} + {{\mathbf{n}}_{{l_s}}}
\end{equation}
where ${{\mathbf{r}}_{{l_s}}}\in {\mathbb{C}}^{{R_{{l_s}}} \times 1}$ denotes the received signal vector of the secondary link ${l_s}$, with ${R_{{l_s}}}$ denoting the numbers of receiving antennas. ${{\mathbf{t}}_{{l_s}}}\in {\mathbb{C}}^{{T_{{l_s}}} \times 1}$ denotes the transmitted signal vector of  link ${l_s}$, with ${T_{{l_s}}}$ denoting the numbers of transmitting antennas.
${{\mathbf{H}}_{{l_s}}} \in {\mathbb{C}^{{R_{{l_s}}} \times {T_{{l_s}}}}}$ denotes the channel gain matrix from the transmitting node to the receiving node of link ${l_s}$.
$\mathbf{H}_{j_pl_s} \in {\mathbb{C}^{{R_{{l_s}}} \times {T_{{j_p}}}}}$
denotes the channel gain matrix from the transmitting node of the primary link $j_p$ to the receiving node of the secondary link ${l_s}$, with ${T_{{j_p}}}$ denoting the number of transmitting antenna of the primary link $j_p$.
$\mathbf{t}_{j_p}$ denotes the transmitted signal vector of the primary link ${j_p}$. ${{\mathbf{n}}_{{l_s}}}$ denotes an additive white Gaussian noise vector of link $l_s$. We assume that the channel is block fading and ${{\mathbf{H}}_{{l_s}}}$ is known at the transmitting node of link ${l_s}$.

\subsection{SU power constraint}

Here, the transmit power constraint of each transmitting node ${n_s}$ of SU link is given by:
\begin{equation}\label{2}
\sum\nolimits_{{l_s} \in \mathcal{O}({n_s})} {\text{Tr}({{\mathbf{Q}}_{{l_s}}})}  \leqslant {\alpha_{n_s} P_{{n_s}}}\;\;\;\forall {n_s}
\end{equation}
where ${{\mathbf{Q}}_{{l_s}}} = {\mathbb{E}}\{ {{\mathbf{t}}_{{l_s}}}{{\mathbf{t}}^H_{{l_s}}}\}$ denotes the transmitting covariance matrix of ${{\mathbf{t}}_{{l_s}}}$ at link ${l_s}$, which is Hermitian and positive semi-definite (PSD).
${P_{{n_s}}}$ denotes the maximum transmitting power  at SU $n_s$.
$0<\alpha_{n_s}\leq1$  is a chosen constant, and denotes the required power reduction with respect to the full power usage at SU $n_s$.  Note that from a mathematical perspective, accounting for $\alpha_{n_s}$ is straightforward.

\subsection{PU Rate Constraint}

In order to ensure the QoS of the PU network, we have the following rate requirement at PU link $j_p$, inspired by\cite{RuiZhangICL2012},
\begin{equation}\label{purequire}
  R_{j_p}\geq R_{j_p}^{\text{min}}
\end{equation}
where ${{{R}}_{{j_p}}^{ \text{min}}}$ denotes the minimum transmission rate requirement of the primary link ${j_p}$.
$R_{j_p}$ denotes the actual transmission rate of PU link ${j_p}$.

We assume that PU link is not fully interfered illustrated in the top-left of Fig. 1.
So, $R_{j_p}$  is composed of two parts\footnote{Note that Eq. \eqref{RpRate} is also suitable for the case where a MIMO SU link shares the full bandwidth with a MIMO PU link.}:
\begin{equation}\label{RpRate}
R_{j_p}=\sum\limits_{{l_s} \in \mathcal{O}({n_s})}W_{{l_s}}r_{j_p}^{l_s}+R_{j_p}^{\text{NI}}
\end{equation}
where ${W_{{l_s}}}\geq0$ denotes the bandwidth assigned to SU link ${l_s}$.
The first term on the RHS is the achievable rate at the frequency band which suffers the interference from SU. Based on the classical capacity formula, we have
\begin{equation}\label{3}
\begin{gathered}
  r_{j_p}^{l_s}=\log \frac{{\left|{{W_{{l_s}}}{N_{{j_p}}}} {{\mathbf{I}} + {\frac{{{W_{{l_s}}}}}{{{B_{{j_p}}}}}} {{\mathbf{H}}_{{j_p}}}{\mathbf{Q}}_{{j_p}}^*{\mathbf{H}}_{{j_p}}^H + {{\mathbf{H}}_{{l_s}{j_p}}}{{\mathbf{Q}}_{{l_s}}}{\mathbf{H}}_{{l_s}{j_p}}^H} \right|}}{{\left|{{N_{{j_p}}}{W_{{l_s}}}}{{\mathbf{I}} + {{\mathbf{H}}_{{l_s}{j_p}}}{{\mathbf{Q}}_{{l_s}}}{\mathbf{H}}_{{l_s}{j_p}}^H} \right|}}
\end{gathered}
\end{equation}
The second term on the RHS $R_{j_p}^{\text{NI}}$ is the achievable rate at the frequency band which does not suffer the interference from SU,
\begin{equation}\label{3B}
\begin{aligned}
&R_{{j_p}}^{{\rm{NI}}} =\\
&\left( {{B_{{j_p}}} - \sum\limits_{{l_s} \in O({n_s})} {{W_{{l_s}}}} } \right){\rm{*}}\log \left| {{\bf{I}} + \frac{1}{{{N_{{j_p}}}{B_{{j_p}}}}}{{\bf{H}}_{{j_p}}}{\bf{Q}}_{{j_p}}^*{\bf{H}}_{{j_p}}^H} \right|
\end{aligned}
\end{equation}
where $N_{{j_p}}$ is the noise power spectral density at PU $j_p$.
${{\mathbf{H}}_{{j_p}}} \in {\mathbb{C}^{{R_{{j_p}}} \times {T_{{j_p}}}}}$ denotes the channel matrix of PU link ${j_p}$ .
${{\mathbf{H}}_{{l_s}{j_p}}} \in {\mathbb{C}^{{R_{{j_p}}} \times {T_{{l_s}}}}}$ denotes the channel matrix from the transmitting node of SU link ${l_s}$ to the transmitting node of PU link ${j_p}$.
We assume based on the active support from the primary network, $R_{j_p}^{\text{min}}$, $B_{{j_p}}$, ${{\mathbf{H}}_{{j_p}}}{\mathbf{Q}}_{{j_p}}^*{\mathbf{H}}_{{j_p}}^H$ and ${{\mathbf{H}}_{{l_s}{j_p}}}$ are known at the transmitting nodes of link $l_s$.

\subsection{Channel Capacity and Bandwidth Allocation}
The capacity of a MIMO link ${l_s}$ is given by
\begin{equation}\label{5_0}
\begin{split}
   & {\Phi _{{l_s}}}({W_{{l_s}}},{{\mathbf{Q}}_{sl}}) \triangleq  \\
    & {W_{{l_s}}}\log \frac{\left|{{W_{{l_s}}}{N_{{l_s}}}}{\mathbf{I}} + {\frac{{{W_{{l_s}}}}}{{{B_{{j_p}}}}}}\mathbf{H}_{j_pl_s}\mathbf{Q}_{j_p}^{*}\mathbf{H}_{j_pl_s}^{H} + {{\mathbf{H}}_{{l_s}}}{{\mathbf{Q}}_{{l_s}}}{\mathbf{H}}_{{l_s}}^H \right|}{\left|{{W_{{l_s}}}{N_{{l_s}}}}\mathbf{I} + {\frac{{{W_{{l_s}}}}}{{{B_{{j_p}}}}}}\mathbf{H}_{j_pl_s}\mathbf{Q}_{j_p}^{*}\mathbf{H}_{j_pl_s}^{H} \right|}
\end{split}
\end{equation}
where $N_{{l_s}}$ is the noise power spectral density at SU $l_s$.
As seen in (\ref{5_0}), the optimization of bandwidth allocation ${W_{{l_s}}}$ and covariance matrix ${{\mathbf{Q}}_{sl}}$ play an important role in improving the channel capacity.
Notice that Eq. \eqref{5_0}  in this paper is different with the capacity function Eq. (2) of a MIMO link used in \cite{LiuJSAC2008},
where the background noise is assumed to be not related to the allocated bandwidth.

Since the total outgoing bandwidth of node ${n_s}$ can not exceed its assigned bandwidth, we have
\begin{equation}\label{5B}
\sum\nolimits_{{l_s} \in \mathcal{O}({n_s})} {{W_{{l_s}}} \leqslant {B_{{n_s}}}}\;\; \forall {n_s}
\end{equation}
${B_{{n_s}}}$ denotes the bandwidth assigned to node ${n_s}$.

 \subsection{Multi-commodity flow model}

 We use a multi-commodity flow model for the routing of data packets in the multihop wireless network. The source nodes send data packets to their intended destination nodes through multi-path and multi-hop routing. We assume there are $F$ sessions in the network. $e_f\geq0$ denotes the traffic demand of session $f$.

 Let $x_{l_s}^f \geqslant 0$ denote the transmission rate of session $f$  over the link ${l_s}$. Following the flow conservation law, we have
 \begin{equation}\label{flowlaw}
 \sum\limits_{{l_s} \in \mathcal{O}({n_s})} {x_{l_s}^f - \sum\limits_{{l_s} \in \mathcal{I}({n_s})} {x_{{l_s}}^f = a_{n_s}e_f} \;\;\;\forall n_s, \forall f}
 \end{equation}
where $\mathcal{I}({n_s})$ denotes the sets of links that are incoming to node ${n_s}$, $a_{n_s}$ is defined as follow

\[{a_{{n_s}}} = \left\{ \begin{array}{l}
 1\;\;\;\text{ if } n_s=\text{src}(f) \\
 -1 \;\;\text{if } n_s=\text{dst}(f)\\
 0 \;\;\;\text{ otherwise}\\
 \end{array} \right.\]
where $\text{src}(f)$ and $\text{dst}(f)$ denote the source and destination node of session $f$, respectively.

Obviously, the total traffic rate of all flows traversing a link cannot exceed the
link's capacity limitation. So we have
\begin{equation}\label{ratecontrol}
\sum\limits_{f = 1}^F {x_{{l_s}}^f \leqslant {\Phi _{{l_s}}}({W_{{l_s}}},{{\mathbf{Q}}_{{l_s}}})\;\;\;\forall l_s}
\end{equation}

Notice that (\ref{ratecontrol}) is a convex constraint due to the concavity of the channel capacity function (\ref{5_0}) in $({W_{{l_s}}},{{\mathbf{Q}}_{{l_s}}})$.
\subsection{Problem Formulation}\label{ProblemFormulation}
Our objective is to optimize the weighted sum of network
utility and power consumption of SU communication
sessions, with a minimum PU transmission rate constraint
and SU power consumption constraints.
We have the following formulation to jointly optimize the covariance matrix $\mathbf{Q} \triangleq [\mathbf{Q}_1,\mathbf{Q}_2,\cdot\cdot\cdot,\mathbf{Q}_{L_s}]$ at the physical layer, bandwidth allocation $\mathbf{W} \triangleq [{W}_1,{W}_2,\cdot\cdot\cdot,{W}_{L_s}]$  at the link layer, routing $\mathbf{x} \triangleq [x_1^1,\cdot\cdot\cdot,x_{L_s}^1,\cdot\cdot\cdot,x_1^F,\cdot\cdot\cdot,x_{L_s}^F]$  at the network layer and rate control $\mathbf{e} \triangleq [e_1,e_2,\cdot\cdot\cdot,e_f]$  at the transport layer:
\begin{eqnarray}\label{6}
    \text{maximize}       &\sum\limits_{f = 1}^F {U_f({e_f}) - \sum\limits_{{l_s} = 1}^{L_s} {{t_{{l_s}}}\text{Tr}({{\mathbf{Q}}_{{l_s}}})} } \\
    \text{subject to}     &(\ref{2}),(\ref{purequire}),(\ref{5B}),(\ref{flowlaw})\text{ and }(\ref{ratecontrol}) \notag
\end{eqnarray}
where $t_{l_s}, \forall {l_s}$ are weighting coefficients to specify the importance of power consumption.

\textbf{Remark 1}:
We extend traditional MIMO CRN design towards a framework of utility-power trade-off with power related objective and constraint. Different from the problem studied in \cite{LiuJSAC2008}, we address the optimization of the weighted sum of network utility and power consumption by introducing the parameter $t_{l_s}$  in \eqref{6},
and consider non-full power consumption by introducing the parameter $\alpha_{n_s}$ in \eqref{2}. We quantify the utility-power trade-off with some surprising results through the simulation verification. Furthermore, the problem is more complex than that in \cite{LiuJSAC2008} due to the non-convex constraint \eqref{purequire} and thus we need to seek a new approach to attack the problem.

\section{Dual Decomposition}
 In this section, we use the dual decomposition technique
to decompose the problem (\ref{6}) into two subproblems. One is the network-transport layer subproblem, i.e., routing at the network layer and rate control at the transport layer.
The other is the physical-link layer subproblem, i.e.,  covariance matrix at the physical layer and bandwidth allocation at the link layer. Both of them can be solved in a distributed fashion.

If only \eqref{ratecontrol} is dualized, the corresponding Lagrangian is
\begin{equation}\label{7}
\begin{gathered}
  L({\mathbf{e}},{\mathbf{x}},{\mathbf{Q,W}},{\mathbf{u}}) \hfill \\
   = \sum\limits_{{n_s} = 1}^{N_s} {\left\{ {\sum\limits_{{l_s} \in \mathcal{O}({n_s})} {\left\{ {{u_{{l_s}}}{\Phi _{{l_s}}}({W_{{l_s}}},{{\mathbf{Q}}_{{l_s}}}) - {t_{{l_s}}}\text{Tr}({{\mathbf{Q}}_{{l_s}}})} \right\}} } \right\}}  \hfill \\
  \;\;\; + \sum\limits_{f = 1}^F {\left\{ {U_f({{e}_f}) - \sum\limits_{{l_s} = 1}^L {u_{l_s}x_{l_s}^f}} \right\}}  \hfill \\
\end{gathered}
\end{equation}
where ${\mathbf{u}} \triangleq [{u_1}, \cdot\cdot\cdot, {u_{l_s}}, \cdot\cdot\cdot ,{u_{L_s}}]$, and $u_{l_s}$ denotes the Lagrange multiplier (i.e., price) associated with the constraint $\sum_{f = 1}^F {x_{{l_s}}^f \leqslant {\Phi _{{l_s}}}({W_{{l_s}}},{{\mathbf{Q}}_{{l_s}}})}$ of link ${l_s}$.

The Lagrange dual function $D({\mathbf{u}})$ is given by
\begin{eqnarray}\label{LagrangeDualFunction}
   D({\mathbf{u}}) \triangleq \text{maximize}       &L({\mathbf{e,x,Q,W,u}}) \notag\\
    \text{subject to}     &(\ref{2}),(\ref{purequire}),(\ref{5B}),(\ref{flowlaw})
\end{eqnarray}

The dual problem is
\begin{equation}\label{DualProblem}
\begin{gathered}
  {\text{minimize}}\;\;\;D({\mathbf{u}}) \hfill \\
  {\text{subject}}\;{\text{to}}\;\;\;{\mathbf{u}} \geqslant {\mathbf{0}} \hfill \\
\end{gathered}
\end{equation}

We notice that the Lagrange (\ref{7}) is separable at each SU node or at each session.
So, we decompose the dual function (\ref{LagrangeDualFunction}) into two classes of subproblems, i.e., the physical-link layer subproblems at each SU node and the network-transport layer subproblems at each session. Specifically, the network-transport layer subproblem at each session is given by
\begin{equation}\label{10}
\begin{gathered}
  {\max_{{{{e}}_f},{{\mathbf{x}}_f}}}\;\;\;\;\;\;\;\;{ {U_f({e_f}) - \sum\limits_{l_{s} = 1}^{L_s} {{u_{{l_s}}}x_{l_s}^f } } }  \hfill \\
  {\text{subject}}\;{\text{to}} \;\;\;\sum\limits_{{l_s} \in \mathcal{O}({n_s})} {x_{l_s}^f - \sum\limits_{{l_s} \in \mathcal{I}({n_s})} {x_{{l_s}}^f = {a_{n_s}e_f}} }\;\;\\
  x_{l_s}^f\geq 0,e_f\geq 0
\end{gathered}
\end{equation}
and the physical-link layer subproblem at each SU node is given by
\begin{equation}\label{11}
\begin{gathered}
  {\max_{{\mathbf{Q}_{n_s}}{\text{,}}{\mathbf{W}_{n_s}}}}\;{
  \sum\limits_{{l_s} \in \mathcal{O}({n_s})} \left({{u_{{l_s}}}{\Phi _{{l_s}}}({W_{{l_s}}},{{\mathbf{Q}}_{{l_s}}})}  -  {{t_{{l_s}}}\text{Tr}({{\mathbf{Q}}_{{l_s}}})}\right)  \hfill
}  \hfill \\
  {\text{subject}}\;{\text{to}}\;\;\;\sum\nolimits_{{l_s} \in \mathcal{O}({n_s})} {{W_{{l_s}}} \leqslant {B_{{n_s}}}} \;\;\; \hfill  \quad\\
   \;\;\;\;\;\;\;\;\;\;\;\;\;\;\;\;\;\;\sum\nolimits_{{l_s} \in \mathcal{O}({n_s})} {\text{Tr}({{\mathbf{Q}}_{{l_s}}})}  \leqslant \alpha_{n_s} {P_{{n_s}}} \hfill \\
  \;\;\;\;\;\;\;\;\;\;\;\;\;\;\;\;\;\;R_{{j_p}}^{min} \leqslant {{ R}_{{j_p}}}\hfill   \quad\\
  \;\;\;\;\;\;\;\;\;\;\;\;\;    {{\mathbf{Q}_{l_s}}}\underset{\raise0.3em\hbox{$\smash{\scriptscriptstyle-}$}}{ \succ } \mathbf{0},{{W}_{l_s}} \geq 0 \;\;\;\;\;\;  \forall {l_s\in \mathcal{O}(n_s)}
\end{gathered}
\end{equation}

We denote $\mathbf{x}_f \triangleq [x_{l_s}^f,\forall l_s]$, $\mathbf{Q}_{n_s} \triangleq [\mathbf{Q}_{l_s},l_s\in\mathcal{O}(n_s)]$, $\mathbf{W}_{n_s} \triangleq [W_{l_s},l_s\in\mathcal{O}(n_s)]$.

\begin{figure*}[t]
\centering
\includegraphics[scale=0.9,bb= 124 664 486 750]{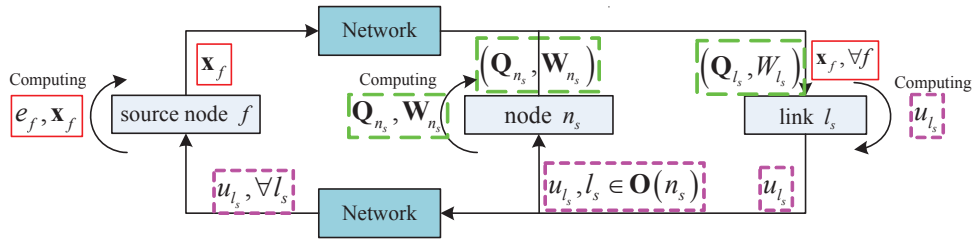}
\caption{Local computation and information exchange}
\label{fig:fig1A}
\end{figure*}

 It is readily seen that the problem (\ref{10}) is convex and can be globally solved at the source node of each session $f$. Although the problem (\ref{11}) can be solved by each node $n_s$, it is difficult to obtain the global solution due to the non-convexity of the constraint $R_{{j_p}}^{min} \leqslant {{ R}_{{j_p}}}$. Hence in the next section we propose an iterative algorithm that converges monotonically.

In the following, we formally solve the dual problem (\ref{DualProblem}). Because the dual objective function is a piece-wise linear function which is non-differentiable, we can use the subgradient method to solve (\ref{DualProblem})\cite{boydcvbook}.
We start with an initial ${{\mathbf{u}}^{(0)}}$. In the $k$-th iteration, after solving the problem (\ref{10}) and the problem (\ref{11}) for given ${{\mathbf{u}}^{(k)}}$, we update the dual variables ${{\mathbf{u}}^{(k + 1)}}$. The $l_s$-th element  $u_{l_s}^{(k + 1)}$ is updated as
\begin{equation}\label{12}
\begin{gathered}
u_{l_s}^{(k + 1)} = {\left[ {{{u_{l_s}^{(k)} - {\beta_{l_s}^{(k)}}{d_{l_s}^{(k)}}}}} \right]^ + }
\end{gathered}
\end{equation}
where $[z]^{+}=\max(z,0)$, ${\beta_{l_s}^{(k)}}>0$ is a positive scalar step size parameter.
The subgradient $d_{l_s}^{(k)}$ can be calculated as
\begin{equation}\label{15}
\begin{gathered}
d_{{l_s}}^{(k)} = {\Phi _{{l_s}}}({{W}_{l_s}^*}({\mathbf{u}^{(k)}}),{{\mathbf{Q}}_{l_s}^*}({\mathbf{u}^{(k)}})) - \sum\limits_{f=1}^F x_{l_s}^{f*}({\mathbf{u}^{(k)}})
\end{gathered}
\end{equation}
where $W_{l_s}^{*}$ and $\mathbf{Q}_{l_s}^{*}$ are
a solution of the problem (\ref{11}) at $n_s$, and $x_{l_s}^{f*}$ is a solution to (\ref{10}) of session $f$ for a given $\mathbf{u}^{(k)}$.

It is worth pointing out that the dual variable $u_{l_s}^{(k)}$ can be interpreted as the ``price'' of link $l_s$ during the $k$-th iteration. This can help us to better understand the update of the dual variables $u_{l_s}^{(k)}$. The subgradient $d_{l_s}^{(k)}$ indicates the usage of the link capacity during the $k$-th iteration. For link $l_s$, when it is under-utilized, then $d_{{l_s}}^{(k)} > 0$, and the price of link $l_s$ will reduce from (\ref{12}). On the other hand, when link $l_s$ is over-utilized, then $d_{{l_s}}^k < 0$, and the price of link $l_s$ will increase.

We now give the following distributed algorithm, where
each session source and each SU solve their own problems with only
local information.

\noindent\rule{520pt}{1pt}
 \textbf{Algorithm 1: Dual decomposition algorithm}\\
 \noindent\rule{520pt}{0.65pt}
Initialization: ${{\mathbf{u}}^{(0)}}$, $\mathbf{Q}^{0}$, $\mathbf{W}^{0}$, $\mathbf{x}^{0}$, $\mathbf{s}^{0}$.\\
Repeat until convergence:\\
 \indent  \indent $k_{1} \leftarrow k_{1}+1$\\
 \indent  \indent At each source node of session $f$, solve $\left({e}_{f}^{(k_{1})},\mathbf{x}_{f}^{(k_{1})}\right)$ \\
  \indent  \indent in the problem (\ref{10}).\\
 \indent  \indent At each node $n_s$, apply Algorithm 2 to solve\\
 \indent  \indent $\left(\mathbf{W}_{l_s}^{(k_{1})},\mathbf{Q}_{l_s}^{(k_{1})}\right),l_s\in\mathcal{O}(n_s)$
 in the problem (\ref{11}) .\\
 \indent  \indent At each node $n_s$, update $u_{l_s}^{(k_{1})}, l_s\in\mathcal{O}(n_s)$ in (\ref{12}).\\
\noindent\rule{520pt}{1pt}

\textbf{Remark 2}: Here, we discuss the distributed implementation of the proposed algorithm.

For the network-transport layer subproblem, after receiving the dual variables, the source node $\text{src}(f)$ of each session $f$ locally solves the network-transport layer subproblem (\ref{10}).
Then the source node has updated the flow rate ${{{s}}_f}$ and routing information ${{\mathbf{x}}_f}$. Each intermediate node performs the routing according to the routing information ${{\mathbf{x}}_f}$ in the source node.

For the physical-link layer subproblem, each node ${n_s}$ locally solves the physical-link layer subproblem (\ref{11}), and then updates ${\mathbf{W}_{{n_s}}}$ and ${{\mathbf{Q}}_{{n_s}}}$.

For the update of the dual variable ${u_{{l_s}}}$ for link ${l_s}$ which is outgoing from node ${n_s}$, from (\ref{12})-(\ref{15}), we notice that the computation of ${u_{{l_s}}}$ only needs the local link capacity information ${\Phi _{{l_s}}}({W_{{l_s}}},{{\mathbf{Q}}_{{l_s}}})$ and the local traffic information $\sum_{f=1}^Fx_{l_s}^f$. Then link ${l_s}$ updates the dual variable. After that, node ${n_s}$ broadcasts the dual information to its next hop neighbors. Meanwhile, it receives the dual information from other neighbors and relays it to its next hop neighbors. Eventually, each source node will get all dual variables of links that the session goes through.

In Fig. \ref{fig:fig1A}, we give a graphic illustration of the local computation and the information exchange.

\section{Linearization-based Alternating Optimization}
In this section, we focus on solving the problem (\ref{11}) through linearization-based alternating optimization. First, it is seen that, fixing $\mathbf{W}_{n_s}$, the problem (\ref{11}) is a nonconvex problem with  respect to $\mathbf{Q}_{n_s}$ due to the primal user rate constraint \eqref{purequire}. While fixing $\mathbf{Q}_{n_s}$, the problem (\ref{11}) is a convex problem with respect to $\mathbf{W}_{n_s}$ but is still not easy to handle. Here, we apply the method of Taylor series
expansion twice to linearize the primal user rate constraint at
each step to get simple convex problem.

Given $\tilde{\mathbf{Q}}_{l_s}$, which satisfies the constraints of the problem \eqref{11},  the rate function \eqref{3} can be linearized as follows\cite{RuiZhangICL2012}:
 \begin{equation}\label{4}
\begin{gathered}
  {\text{ }}{r_{{j_p}}^{l_s}}{\text{ = }}\log \frac{{\left| {{{W_{{l_s}}}{N_{{j_p}}}}{\mathbf{I}} + {\frac{{{W_{{l_s}}}}}{{{B_{{j_p}}}}}} {{\mathbf{H}}_{{j_p}}}{\mathbf{Q}}_{{j_p}}^*{\mathbf{H}}_{{j_p}}^H + {{\mathbf{H}}_{{l_s}{j_p}}}{{\mathbf{Q}}_{{l_s}}}{\mathbf{H}}_{{l_s}{j_p}}^H} \right|}}{{\left| {{{W_{{l_s}}}{N_{{j_p}}}}{\mathbf{I}} + {{\mathbf{H}}_{{l_s}{j_p}}}{{\mathbf{Q}}_{{l_s}}}{\mathbf{H}}_{{l_s}{j_p}}^H} \right|}} \hfill \\
  \;\;\;\;\;\;\; \simeq \log \left| {{{W_{{l_s}}}{N_{{j_p}}}}{\mathbf{I}} + {\frac{{{W_{{l_s}}}}}{{{B_{{j_p}}}}}} {{\mathbf{H}}_{{j_p}}}{\mathbf{Q}}_{{j_p}}^*{\mathbf{H}}_{{j_p}}^H + {{\mathbf{H}}_{{l_s}{j_p}}}{{\mathbf{Q}}_{{l_s}}}{\mathbf{H}}_{{l_s}{j_p}}^H} \right| \hfill \\
  \;\;\;\;\;\;\; - \log \left| {{{W_{{l_s}}}{N_{{j_p}}}}{\mathbf{I}} + {{\mathbf{H}}_{{l_s}{j_p}}}{{{\mathbf{\tilde Q}}}_{{l_s}}}{\mathbf{H}}_{{l_s}{j_p}}^H} \right| \hfill \\
  \;\;\;\;\;\;\; - \text{Tr}\left\{ {{{\left( {{{W_{{l_s}}}{N_{{j_p}}}}{\mathbf{I}} + {{\mathbf{H}}_{{l_s}{j_p}}}{{{\mathbf{\tilde Q}}}_{{l_s}}}{\mathbf{H}}_{{l_s}{j_p}}^H} \right)}^{ - 1}}{{\mathbf{H}}_{{l_s}{j_p}}}{{\mathbf{Q}}_{{l_s}}}{\mathbf{H}}_{{l_s}{j_p}}^H} \right\} \hfill \\
  \;\;\;\;\;\;\; + \text{Tr}\left\{ {{{\left( {{{W_{{l_s}}}{N_{{j_p}}}}{\mathbf{I}} + {{\mathbf{H}}_{{l_s}{j_p}}}{{{\mathbf{\tilde Q}}}_{{l_s}}}{\mathbf{H}}_{{l_s}{j_p}}^H} \right)}^{ - 1}}{{\mathbf{H}}_{{l_s}{j_p}}}{{{\mathbf{\tilde Q}}}_{{l_s}}}{\mathbf{H}}_{{l_s}{j_p}}^H} \right\} \hfill \\
  \;\;\;\;\;\;\;\triangleq \tilde{r}_{{j_pQ}}^{l_s} \hfill \\
\end{gathered}
\end{equation}

So the rate of PU link \eqref{RpRate} can be approximated as
\begin{equation}\label{LinearizedRate}
R_{j_p}\simeq \sum \limits_{l_s \in \mathcal{O}(n_s)}W_{{l_s}}\tilde{r}_{j_pQ}^{l_s}+R_{j_p}^{\text{NI}}\triangleq \tilde{R}_{j_pQ}
\end{equation}

After fixing $\mathbf{W}_{n_s}$ to $\mathbf{\tilde W}_{n_s} = [\tilde W_{l_s},l_s\in\mathcal{O}(n_s)]$, using this linearized rate function ${{\tilde R}_{{j_pQ}}}$, the problem \eqref{11} is transformed into the problem \eqref{11-2}, which is convex in ${\bf{Q}}_{{n_s}}$.
\begin{equation}\label{11-2}
\begin{gathered}
{\max _{{{\bf{Q}}_{{n_s}}}}}\;\;\;\sum\limits_{{l_s} \in {\cal O}\left( {{n_s}} \right)} {\left( {{u_{{l_s}}}{\Phi _{{l_s}}}\left( {{{{{\tilde W}}}_{{l_s}}},{{\bf{Q}}_{{l_s}}}} \right) - {t_{{l_s}}}{\rm{Tr}}\left( {{{\bf{Q}}_{{l_s}}}} \right)} \right)} \\
{\rm{subject}}\;{\rm{to}}\;\;\;\sum\nolimits_{{l_s} \in {\cal O}\left( {{n_s}} \right)} {{\rm{Tr}}({{\bf{Q}}_{{l_s}}})}  \le {\alpha _{{n_s}}}{P_{{n_s}}}\\
R_{{j_p}}^{min} \le {{ \tilde{R}}_{{j_pQ}}}\\
\;\;\;\;\;\;\;\;\;\;{{\bf{Q}}_{{l_s}}} \succ {\bf{0}}\;\;\;\;\;\forall {l_s} \in {\cal O}\left( {{n_s}} \right)
\end{gathered}
\end{equation}
For given $\mathbf{\tilde W}_{n_s}$,  it can be readily verified that
${\Phi _{{l_s}}}\left( {{{{{\tilde W}}}_{{l_s}}},{{\bf{Q}}_{{l_s}}}} \right)$ is concave with respect to
 ${{\bf{Q}}_{{l_s}}}$. All constraints of the problem \eqref{11-2} are linear with respect to
 ${{\bf{Q}}_{{n_s}}}$. Thus, the problem \eqref{11-2} is a convex optimization problem with respect to
 ${{\bf{Q}}_{{n_s}}}$.

Given $\tilde{\mathbf{W}}_{l_s}$, which satisfies the constraints of the problem \eqref{11},  ${W_{{l_s}}}r_{{j_p}}^{{l_s}}$ can be linearized as follows:

 \begin{equation}\label{4}
\begin{array}{l}
{W_{{l_s}}}r_{{j_p}}^{{l_s}}
 = {W_{{l_s}}}\log \frac{{\left| {{N_{{j_p}}}{W_{{l_s}}}{\bf{I}} + \frac{{{W_{{l_s}}}}}{{{B_{{j_p}}}}}{{\bf{H}}_{{j_p}}}{\bf{Q}}_{{j_p}}^*{\bf{H}}_{{j_p}}^H + {{\bf{H}}_{{l_s}{j_p}}}{{{\rm{\tilde Q}}}_{{l_s}}}{\bf{H}}_{{l_s}{j_p}}^H} \right|}}{{\left| {{N_{{j_p}}}{W_{{l_s}}}{\bf{I}} + {{\bf{H}}_{{l_s}{j_p}}}{{{\rm{\tilde Q}}}_{{l_s}}}{\bf{H}}_{{l_s}{j_p}}^H} \right|}}\\
 \simeq {W_{{l_s}}}\log \left| {{N_{{j_p}}}{{\tilde W}_{{l_s}}}{\bf{I}} + \frac{{{{\tilde W}_{{l_s}}}}}{{{B_{{j_p}}}}}{{\bf{H}}_{{j_p}}}{\bf{Q}}_{{j_p}}^*{\bf{H}}_{{j_p}}^H + {{\bf{H}}_{{l_s}{j_p}}}{{{\rm{\tilde Q}}}_{{l_s}}}{\bf{H}}_{{l_s}{j_p}}^H} \right|\\
\;\;\; + {\rm{Tr\{ }}{\left( {{N_{{j_p}}}{{\tilde W}_{{l_s}}}{\bf{I}} + \frac{{{{\tilde W}_{{l_s}}}}}{{{B_{{j_p}}}}}{{\bf{H}}_{{j_p}}}{\bf{Q}}_{{j_p}}^*{\bf{H}}_{{j_p}}^H + {{\bf{H}}_{{l_s}{j_p}}}{{{\bf{\tilde Q}}}_{{l_s}}}{\bf{H}}_{{l_s}{j_p}}^H} \right)^{ - 1}}*\\
{\kern 1pt} {\kern 1pt} \;\;\;\left( {{N_{{j_p}}}{\rm{ + }}\frac{{{{\bf{H}}_{{j_p}{l_s}}}{\bf{Q}}_{{j_p}}^*{\bf{H}}_{{j_p}{l_s}}^H}}{{{B_{{j_p}}}}}} \right){\rm{\} }}\left( {{W_{{l_s}}} - {{\tilde W}_{{l_s}}}} \right){{\tilde W}_{{l_s}}}\\
\;\;\; - {W_{{l_s}}}\log \left| {{N_{{j_p}}}{W_{{l_s}}}{\bf{I}} + {{\bf{H}}_{{l_s}{j_p}}}{{{\bf{\tilde Q}}}_{{l_s}}}{\bf{H}}_{{l_s}{j_p}}^H} \right|\;\;\;
\end{array}
\end{equation}
So the rate of PU link \eqref{RpRate} can be approximated as
\begin{equation}\label{LinearizedRate}
R_{j_p}\simeq \sum \limits_{l_s \in \mathcal{O}(n_s)}\tilde{r}_{j_pW}^{l_s}+R_{j_p}^{\text{NI}}\triangleq \tilde{R}_{j_pW}
\end{equation}

After fixing $\mathbf{Q}_{n_s}$ to ${{{\mathbf{\tilde Q}}}_{{l_s}}}$, using this linearized rate function ${{\tilde R}_{{j_pW}}}$, the problem \eqref{11} is transformed into
the problem \eqref{11-3}, which is convex in ${\bf{W}}_{{n_s}}$.
\begin{equation}\label{11-3}
\begin{gathered}
{\max _{{{\bf{W}}_{{n_s}}}}}\;\;\;\sum\limits_{{l_s} \in {\cal O}\left( {{n_s}} \right)} {\left( {{u_{{l_s}}}{\Phi _{{l_s}}}\left( {{{{{W}}}_{{l_s}}},{{{\mathbf{\tilde Q}}}_{{l_s}}}} \right) - {t_{{l_s}}}{\rm{Tr}}\left( {{{\mathbf{\tilde Q}}_{{l_s}}}} \right)} \right)} \\
{\rm{subject}}\;{\rm{to}}\;\;\;\sum\nolimits_{{l_s} \in {\cal O}\left( {{n_s}} \right)} {{W_{{l_s}}} \le {B_{{n_s}}}} \\
\;\;\;\;\;\;\;\;\;\;\;\;\;\;\;R_{{j_p}}^{min} \le {{ \tilde{R}}_{{j_pW}}}\\
\;\;\;\;\;\;\;\;\;\;\;\;\;\;\;{W_{{l_s}}} \ge 0\;\;\;\;\;\;\forall {l_s} \in {\cal O}\left( {{n_s}} \right)\;\;\;
\end{gathered}
\end{equation}

For given $\mathbf{\tilde Q}_{n_s}$,  it can be readily verified that
${\Phi _{{l_s}}}\left( {{{{{W}}}_{{l_s}}},{{\bf{\tilde Q}}_{{l_s}}}} \right)$ is concave with respect to
 ${{W}_{{l_s}}}$. All constraints of the problem \eqref{11-3} are linear with respect to
 ${{\bf{W}}_{{n_s}}}$. Thus, the problem \eqref{11-3} is a convex optimization problem with respect to
 ${{\bf{W}}_{{n_s}}}$.

 Let $\mathbf{Q}_{n_s}^*$ be its solution to the problem \eqref{11-2} and $f(\mathbf{W}_{n_s}, \mathbf{Q}_{n_s})$ denotes the objective value of the problem \eqref{11-2} or the problem \eqref{11-3}. We have $f(\mathbf{W}_{n_s}, \mathbf{Q}_{n_s}^*)\geq f(\mathbf{W}_{n_s}, \tilde{\mathbf{Q}}_{n_s})$ since $\tilde{\mathbf{Q}}_{n_s}$ is a feasible solution to the problem \eqref{11-2}. We then fix $\mathbf{Q}_{n_s}$ to $\mathbf{Q}_{n_s}^*$ in the problem \eqref{11-3} and solve for $\mathbf{W}_{n_s}$ in \eqref{11-3}. Let $\mathbf{W}_{n_s}^*$ be its solution to the problem \eqref{11-3}. Then we have $f(\mathbf{W}_{n_s}^*, \mathbf{Q}_{n_s}^*)\geq f(\mathbf{W}_{n_s}, \mathbf{Q}_{n_s}^*)$. By the above analysis, we propose Algorithm 2 for solving problem \eqref{11}.
As Algorithm 2 iterates, the objective value of the problem \eqref{11} is nondecreasing.

\noindent\rule{520pt}{1pt}
 \textbf{Algorithm 2: Linearization-based Alternating Optimization Algorithm}\\
 \noindent\rule{520pt}{0.65pt}
Initialization: $k_2=0$, ${\mathbf{W}}_{{n_s}}^{(0)}$, ${\mathbf{Q}}_{{n_s}}^{(0)}$.\\
Repeat until convergence:\\
 \indent  \indent $k_{2} \leftarrow k_{2}+1$;\\
\indent \indent  Linearize $R_{j_p}$ at $\mathbf{Q}_{n_s}^{(k_{2})}$s to get $\tilde{R}_{j_pQ}$;\\
\indent \indent  Fix ${{\mathbf{W}}_{{n_s}}}= {\mathbf{W}}_{{n_s}}^{(k_{2})}$, solve \eqref{11-2} to get ${\mathbf{Q}}_{{n_s}}^{(k_{2} + 1)}$;\\
\indent \indent  Linearize $R_{j_p}$ at $\mathbf{W}_{n_s}^{(k_{2})}$s to get $\tilde{R}_{j_pW}$;\\
\indent \indent  Fix ${{\mathbf{Q}}_{{n_s}}} = {\mathbf{Q}}_{{n_s}}^{(k_{2} + 1)}$, solve \eqref{11-3} to get ${\mathbf{W}}_{{n_s}}^{(k_{2}+1)}$.\\
 \noindent\rule{520pt}{1pt}

\section{Recovery of Primal Solution }
Because of the non-convexity of the primal problem \eqref{6}, the duality gap usually exists and the solution to the dual problem $D({\mathbf{u}})$ is not always feasible for the primal problem \eqref{6}. Hence, we still need to generate a feasible primal solution with good performance. Our key idea is to fix $\mathbf{W}$ in the primal problem. Although the resultant problem is still nonconvex, we can iteratively linearize the primal rate constraint to get a sequence of convex problems that can be distributively solved by the dual-decomposition method. Such an iterative algorithm is called constrained concave-convex procedure (CCCP)\cite{CCCP_convergence} which can monotonically converge to a stationary point of the primal problem with fixed bandwidth allocation.

To save the communication overhead and considering that the dual decomposition algorithm can quickly converge to the neighborhood of the optimal solution (generally oscillating around the optimal solution), we run a few iterations of the dual decomposition algorithm and perform a heuristic step, i.e., average $W_{l_s}$ over iterations,
\begin{equation}\label{heuristic}
{{\mathbf{\bar W}}^*} = \frac{1}{n}{{\sum\limits_{k = N - n + 1}^N {{{\mathbf{W}}^{(k)}}} }}
\end{equation}
where $N$ represents the total number of iterations and $n$ represents the last $n$ iterations. Intuitively, we consider that ${{\mathbf{\bar W}}^*}$ is the neighborhood of the optimal solution ${\mathbf{W}}$. Hence, we fix $\mathbf{W}=\bar{\mathbf{W}}^*$ and linearize the primal user rate function $R_{j_p}$ at ${\mathbf{\tilde Q}}$ to get a convex problem. Then we can use the dual-decomposition method to solve the resultant problem in a distributed fashion. Finally, we get a recovery solution. The whole algorithm is summarized in Algorithm 3. Note that, in Algorithm 3, ${\mathbf{\tilde Q}}$ is initialized from the result obtained by Algorithm 1 which is feasible to the primal problem. Hence, the recovery solution is also feasible for the primal problem\cite{CCCP_convergence}.

\noindent\rule{520pt}{1pt}
 \textbf{Algorithm 3: Primal solution recovery algorithm}\\
 \noindent\rule{520pt}{0.65pt}
Implement Algorithm 1.\\
Get ${\mathbf{\bar W}}^*$ in (\ref{heuristic}).\\
Choose ${{\mathbf{u}}^{(0)}}$ and initialize $\tilde{\mathbf{Q}}$ from the result obtained from Algorithm 1.\\
Repeat until the required accuracy:\\
\indent  \indent  Let $k_3=0$. \\
\indent  \indent  Repeat until the required accuracy:\\
\indent  \indent \indent \indent $k_{3} \leftarrow k_{3}+1$\\
\indent  \indent \indent \indent At each source node of session $f$, compute\\
 \indent  \indent \indent \indent $\left({s}_{f}^{(k_{3})},\mathbf{x}_{f}^{(k_{3})}\right)$ by solving the problem (\ref{10}).\\
 \indent  \indent \indent  \indent  At each node $n_s$, compute $\mathbf{Q}_{l_s}^{(k_{3})},l_s\in\mathcal{O}(n_s)$
 by \\
 \indent  \indent \indent \indent solving the problem (\ref{11}) with ${\mathbf{\bar W}}^*$.\\
 \indent  \indent \indent  \indent  At each node $n_s$, update $u_{l_s}^{(k_{3})}, l_s\in\mathcal{O}(n_s)$ in (\ref{12}).\\
\indent \indent Update ${\mathbf{\tilde Q}} = {{\mathbf{Q}}^{(k_3)}}$, linearize PU rate in (\ref{LinearizedRate}), and update\\
\indent  \indent ${{\mathbf{u}}^{(0)}}={{\mathbf{u}}^{(k_3)}}$.\\
 \noindent\rule{520pt}{1pt}

\section{SIMULATION RESULTS}
\begin{figure}
\centering
\includegraphics[scale=0.5,width=0.5\textwidth]{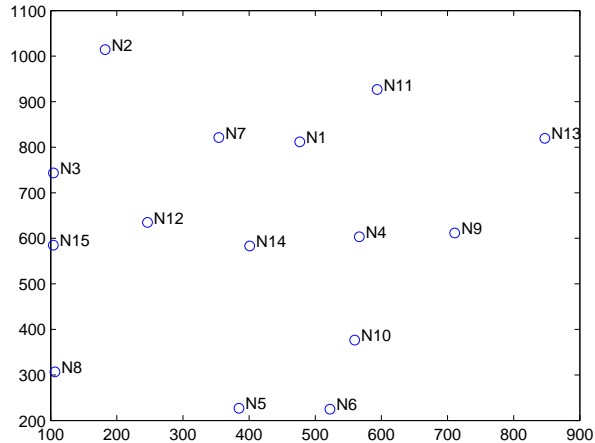}
\centering
\caption{Topology of SU network with 15 nodes.}
\centering
\label{fig:fig1}
\centering
\end{figure}

In this section, we investigate the performance of the proposed distributed algorithm through simulations. The proportional fairness utility function is adopted, i.e., $U_f(f)=\ln (e_f)$ for each session $f$. We randomly generate a SU network topology as shown in the Fig.\ref{fig:fig1}. In this network topology, there are 15 nodes, 56 links\footnote{We assume that there exists an SU link if the distance between two SU nodes is not larger than 300m.}, and three sessions: node 2 to node 6, node 8 to node 13, and node 15 to node 9. Each node in the network is equipped with two transmitting antennas and two receiving antennas, and the allocated bandwidth is 20. The maximum transmit power of each node is 100. The channel gain matrix from the transmitting node of link $l_s$ to the receiving node of
link $l_s$ is modeled as
\begin{equation}\label{ChanelModel}
\mathbf{H}_{l_s}=(200/d_{l_s})^{3.5}L_{l_s}\bar{\mathbf{H}}_{l_s}
\end{equation}
where $d_{l_s}$ is the distance from the transmitting node of link $l_s$ to the receiving node of link $l_s$; $10\log_{10}(L_{l_s})$ is
a real Gaussian random variable with zero mean and a standard deviation of 8 accounting for
the large scale
log-normal shadowing; finally, $\bar{\mathbf{H}}_{l_s}$ is an 2-by-2 matrix containing random values drawn from the standard normal distribution.
Also, the channel gain matrix of PU link and the interference channel gain matrix from the transmitting node of SU link ${l_s}$ to the receiving node of PU link ${j_p}$ are
nearly similar as (\ref{ChanelModel}) with different distance. Specifically, we set the distance of PU link ${j_p}$ as $d_{j_p}=200m$, and the distance from the transmitting node of SU link ${l_s}$ to the receiving node of PU link ${j_p}$ as $d_{s_lp_j}=300m$. We assume that SU $n_s$ shares the full bandwidth with PU link $j_p$. The transmitting covariance matrix $\mathbf{Q}_{j_p}^{*}$ of PU link $j_p$ is determined by the optimal value under the condition of interference from the SU links.
We denote $\rho_{j_p}=R_{j_p}^{\text{min}}/\overline{R_{j_p}}$,
where $\overline{R_{j_p}}$ denotes the maximal rate of PU $j_p$ achieved without any interferences from SUs.
We set the default value of $\rho_{j_p}$ for all PUs as 0.5.

\begin{figure}
\centering
\includegraphics[scale=0.5,width=0.5\textwidth]{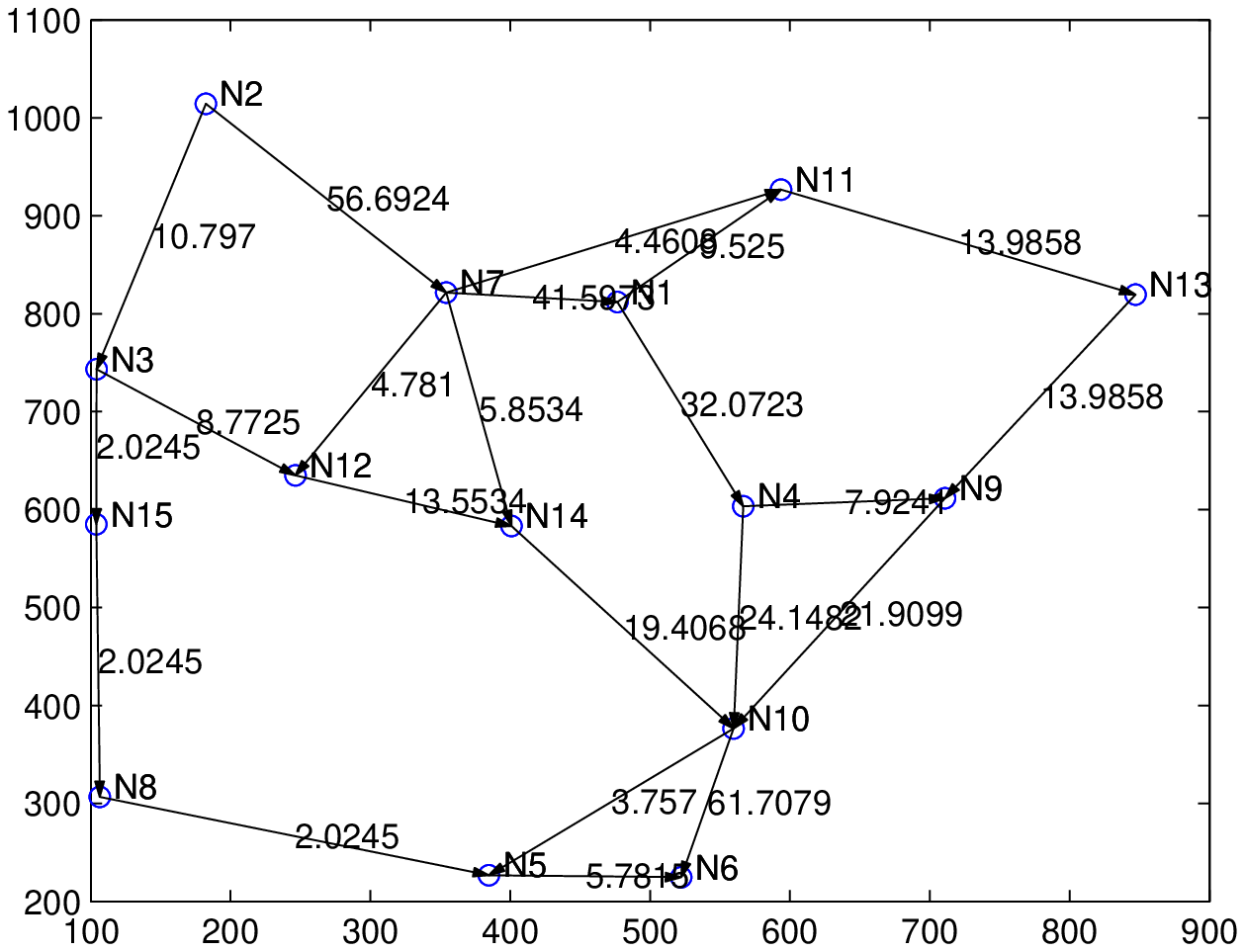}
\centering
\caption{Routing and flow rates of session 1.}
\centering
\label{fig:fig2}
\centering
\end{figure}

\begin{figure}
\centering
\includegraphics[scale=0.5,width=0.5\textwidth]{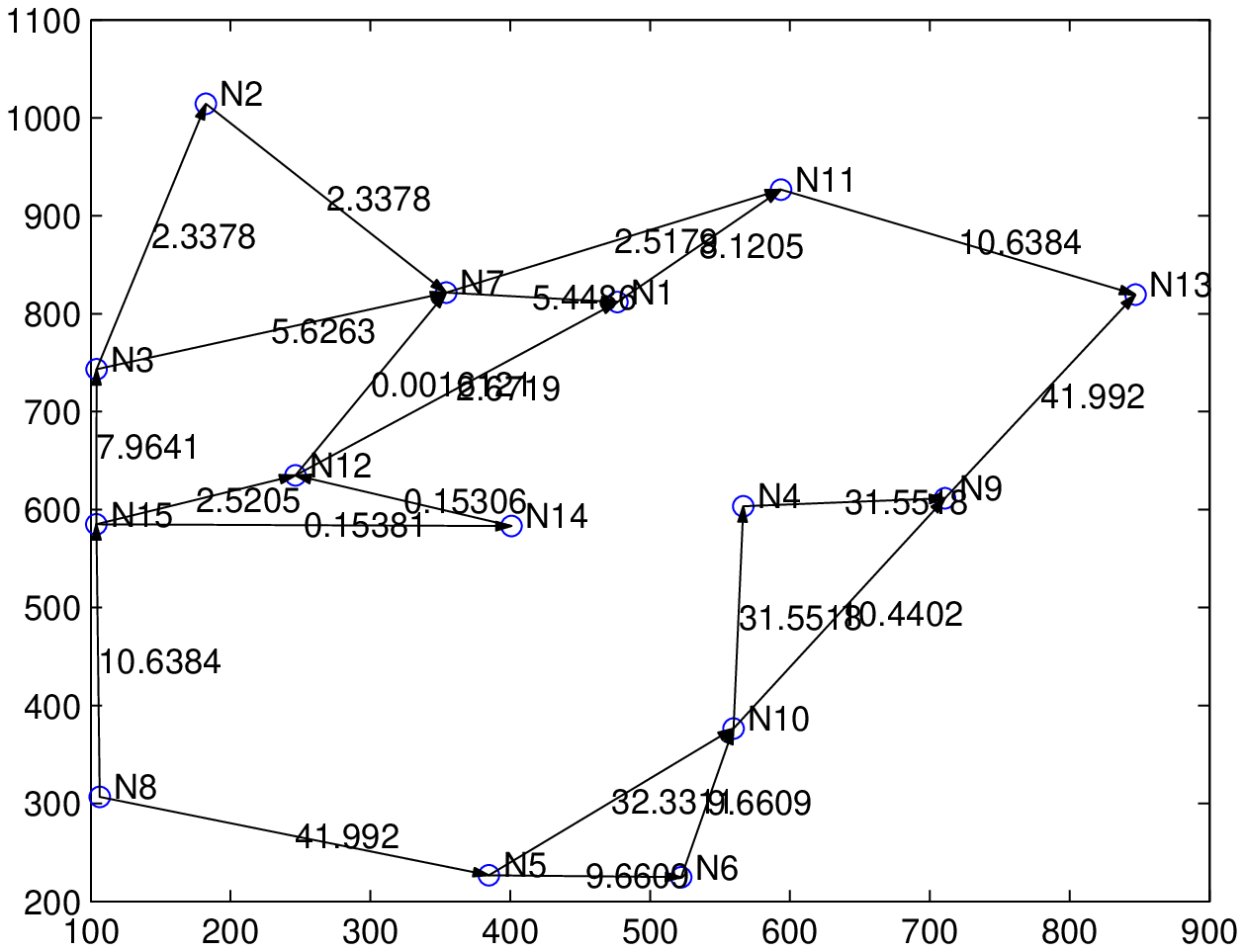}
\centering
\caption{Routing and flow rates of session 2.}
\centering
\label{fig:fig3}
\centering
\end{figure}

\begin{figure}
\centering
\includegraphics[scale=0.5,width=0.5\textwidth]{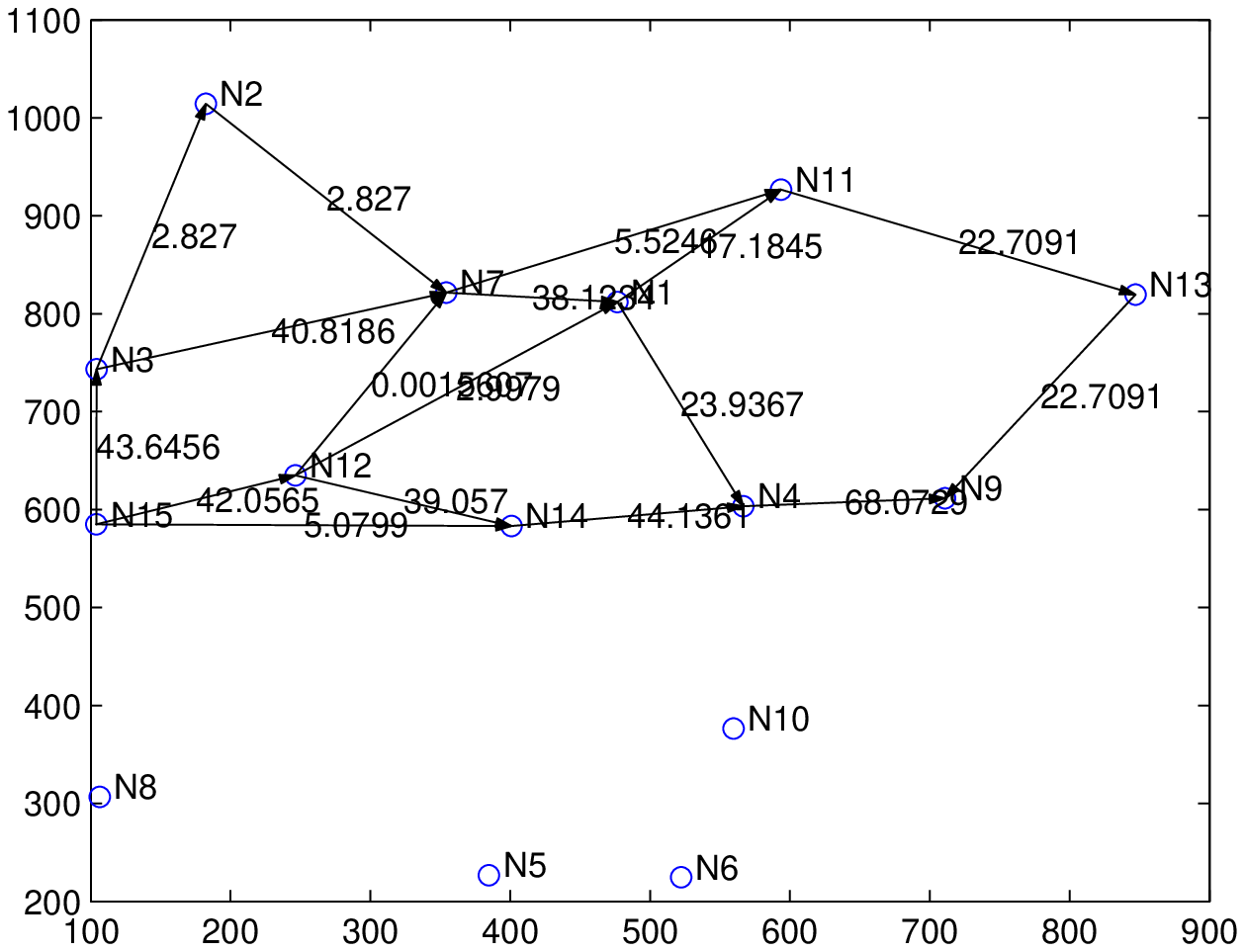}
\centering
\caption{Routing and flow rates of session 3 .}
\centering
\label{fig:fig4}
\centering
\end{figure}

\begin{figure}
\centering
\includegraphics[scale=0.5,width=0.5\textwidth]{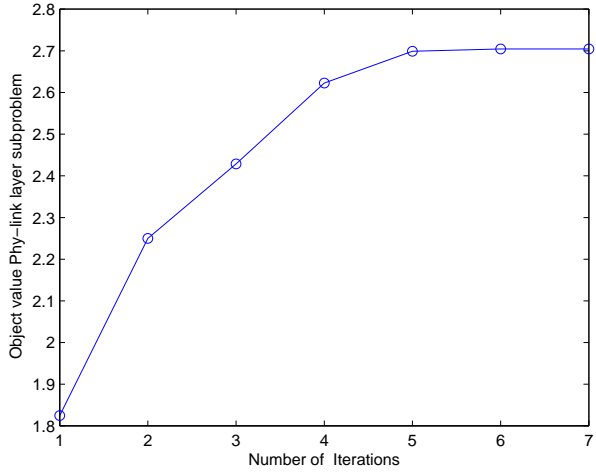}
\centering
\caption{Convergence behavior of linearization-based alternative algorithm for the physical-link layer subproblem.}
\centering
\label{fig:fig5}
\centering
\end{figure}

\begin{figure}
\centering
\includegraphics[scale=0.5,width=0.5\textwidth]{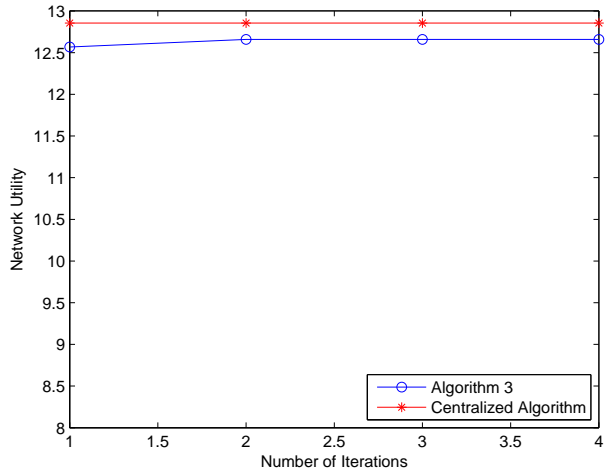}
\centering
\caption{Convergence of the outer loop in Algorithm 3 with $n=20$.}
\centering
\label{fig:fig6}
\centering
\end{figure}

\begin{figure}
\centering
\includegraphics[scale=0.5,width=0.5\textwidth]{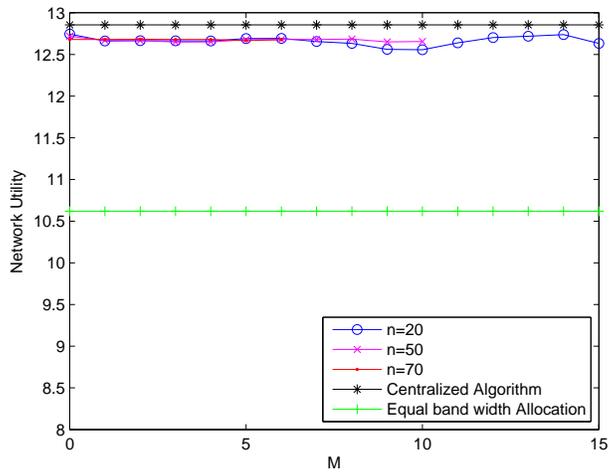}
\centering
\caption{Recovery of primal solution with $n=20,50,70$. (Compared to the centralized algorithm and the equal bandwidth allocation algorithm)}
\centering
\label{fig:fig7}
\centering
\end{figure}

\begin{figure}
\centering
\includegraphics[scale=0.5,width=0.5\textwidth]{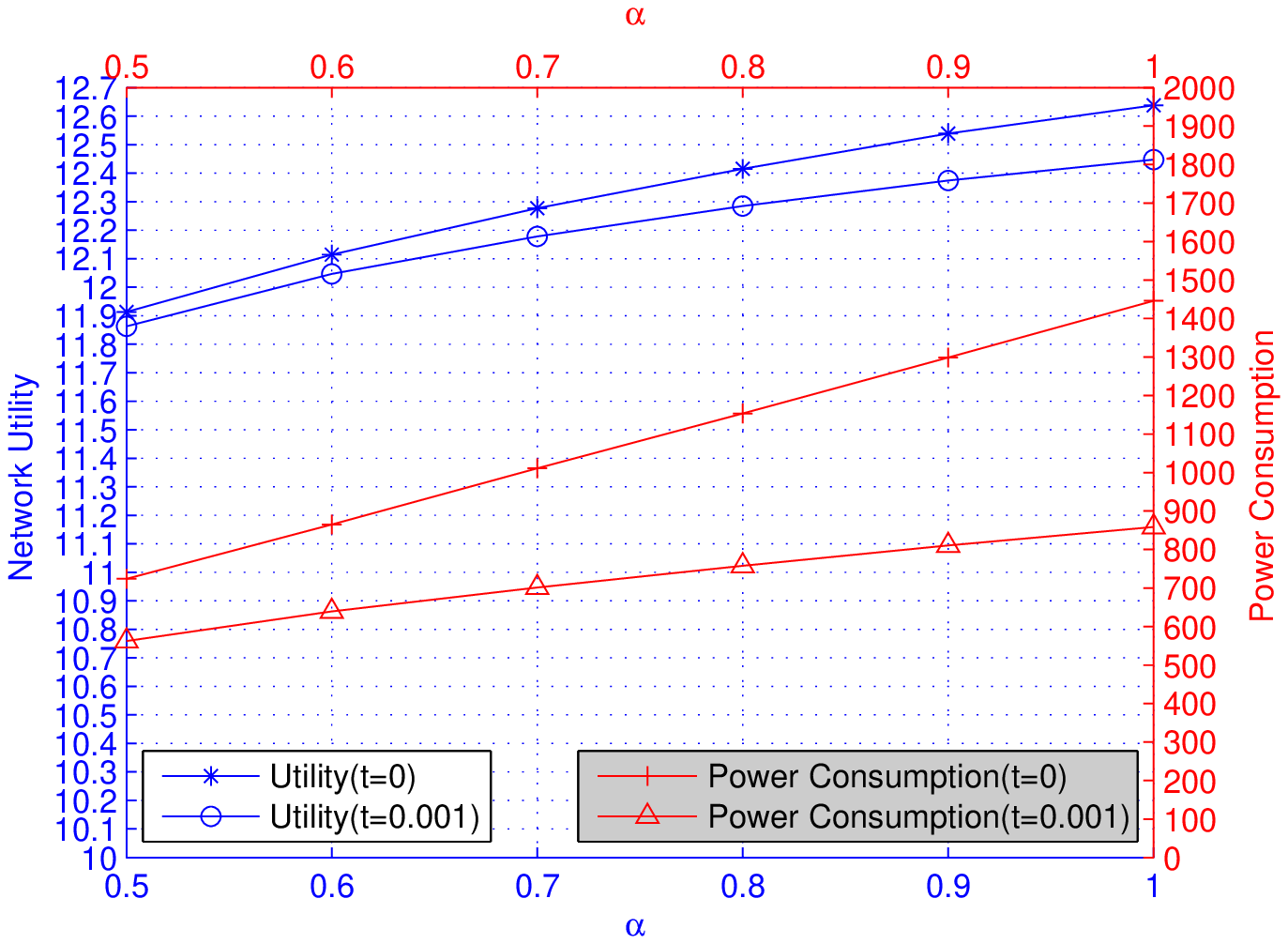}
\centering
\caption{Impact of $\alpha$ and $t$ on network utility and power consumption.}
\centering
\label{fig:fig8}
\centering
\end{figure}

Through implementing the proposed distributed algorithm, we get the routing and flow rates of sessions 1, 2 and 3, which are shown in Figs. \ref{fig:fig2},\ref{fig:fig3} and \ref{fig:fig4}, respectively. It is seen that flow routings for sessions 1, 2 and 3 are multihop and multi-path. Moreover, it can be easily verified that the flow rates satisfy the flow conservation.

Fig. \ref{fig:fig5} shows the convergence behavior of the linearization-based alternating optimization algorithm for the physical-link layer subproblem with fixed $\mathbf{u}$, while Fig. \ref{fig:fig6} shows the convergence behavior of the recovery algorithm with fixed bandwidth allocation. We can see that, the two algorithms can converge quickly and keep the objective function nondecreasing at each iteration. Particularly, the recovery algorithm can achieve the most of performance in the first iteration. Hence, to save the communication overhead, we only need to implement once the dual decomposition method in the recovery step.

To examine the performance of the bandwidth-allocation scheme, we run the dual-decomposition algorithm 300 iterations in total. We set $n=20,50,70$ and Let $N= 200+n+5M$, $M=0,1,2,\ldots,15$. For each combination $(n, N)$, we carry out the recovery step and calculate the network utility. Fig. \ref{fig:fig7} shows the simulation results. We can see that, for different SU $n_s$ , the proposed algorithm has very similar performance. Moreover, the algorithm is also robust to the choice of $N$, i.e., the total number of iterations in the first step. Hence, in the other simulations, we set $N=250$ and $n=20$.

From Fig. \ref{fig:fig7}, we can also see that the proposed bandwidth allocation scheme is better than the equal bandwidth allocation scheme. Moreover, the performance of the proposed distributed algorithm is very close to the performance of the centralized algorithm using Matlab solver \emph{fmincon}. It is worth mentioning that, although \emph{fmincon} can only offer locally optimal solution, we find from simulations that, it can generate almost the same utility from different initialization. Hence, our distributed algorithm can find a feasible solution with good performance.

We investigate the impact of two parameters $\alpha_{n_s}$ and $t_{l_s}$ on the achieved network utility and the power consumption. The simulation result is shown Fig. \ref{fig:fig8}, where $\alpha_{n_s}=\alpha, \forall n_s$ is chosen from 0.5 to 1, and $t_{l_s}=t, \forall l_s$  is chosen as 0 and 0.001, respectively. It is seen that there exist a good tradeoff between the network utility and the power consumption. Through choosing suitable $\alpha_{n_s}$ and $t_{l_s}$, we can significantly save the power consumption, while achieving the network utility very close to that achieved at the full power consumption. For example, shown in Fig. \ref{fig:fig8}, when $\alpha=0.7$ and $t=0.001$,
we can achieve up to 99\% of the network utility achieved at full power consumption, while only expending about $70$ of full power. Thus, our formulation provides an efficient way to design a green cross-layer optimization scheme for multihop MIMO CRN.

\section{Conclusions}
Both MIMO and CR are the enabling technologies for the next-generation wireless communications.
The performance of multihop MIMO CRN is tightly coupled with mechanisms at the physical, link, network, and transport layers. In this paper, we have proposed a new formulation for green multihop MIMO CRN design. Our formulation balances the network utility and weighted total power consumption of SU communication sessions, with a minimum PU transmission rate constraint and SU power constraints. We have developed the distributed algorithms to tackle the highly nonconvex cross-layer optimization problem. Simulation results show that the proposed algorithm can provide a feasible solution with good performance, and the proposed formulation is power-efficient while maintaining the high network utility.

\ifCLASSOPTIONcaptionsoff
  \newpage
\fi

\end{document}